\documentclass[referee]{raa}
\usepackage{graphicx,times}
\usepackage{natbib}
\usepackage{amssymb,amsmath}
\bibpunct{(}{)}{;}{a}{}{,}

\usepackage[a4paper=true,dvipdfm=true,pagebackref=true]{hyperref}
\hypersetup{pdftitle = The title of my PDF, pdfauthor = My name, pdfsubject= The subject, pdfkeywords = keyword1 keyword2 keyword3}
\hypersetup{colorlinks = true, linkcolor = green, anchorcolor = red, citecolor = blue, filecolor = red, pagecolor = red, urlcolor = red}

\begin{document}

   \title{New Limits on the Photon Mass with Radio Pulsars in the Magellanic Clouds}

 \volnopage{ {\bf 2016} Vol.\ {\bf X} No. {\bf XX}, 000--000}
   \setcounter{page}{1}

   \author{Jun-Jie Wei\inst{1,2}, Er-Kang Zhang\inst{1}, Song-Bo Zhang\inst{1} and Xue-Feng Wu\inst{1,3}
   }

   \institute{Purple Mountain Observatory, Chinese Academy of Sciences, Nanjing 210008, China; {\it jjwei@pmo.ac.cn; xfwu@pmo.ac.cn}\\
	\and Guangxi Key Laboratory for Relativistic Astrophysics, Nanning 530004, China\\
	\and Joint Center for Particle, Nuclear Physics and Cosmology, Nanjing University-Purple Mountain Observatory, Nanjing 210008, China\\
\vs \no
   {\small Received----- ; accepted -----}
}

\abstract{A conservative constraint on the rest mass of the photon can be estimated under
the assumption that the frequency dependence of dispersion from astronomical sources is
mainly contributed by the nonzero photon mass effect. Photon mass limits have been earlier
set through the optical emissions of the Crab Nebula pulsar, but we prove that these limits
can be significantly improved with the dispersion measure (DM) measurements of radio pulsars
in the Large and Small Magellanic Clouds. The combination of DM measurements of pulsars and
distances of the Magellanic Clouds provide a strict upper limit on the photon mass as low as
$m_{\gamma} \leq2.0\times10^{-45}~\rm{g}$, which is at least four orders of magnitude smaller
than the constraint from the Crab Nebula pulsar. Although our limit is not as tight as
the current best result ($\sim10^{-47}~\rm{g}$) from a fast radio burst (FRB 150418)
at a cosmological distance, the cosmological origin of FRB 150418 remains under debate;
and our limit can reach the same high precision of FRB 150418 when it has an extragalactic origin
($\sim10^{-45}~\rm{g}$).
}

   \authorrunning{J. J. Wei et al. }            
   \titlerunning{Photon Mass limits from Radio Pulsars}  
   \maketitle

%
\section{Introduction}           
\label{sect:intro}

The cornerstones of modern physics, such as the classical Maxwellian electromagnetism and
the second postulate of Einstein's theory of special relativity, have a basic assumption
that all electromagnetic radiation travels in vacuum at the constant speed $c$, which implies
that the photon rest mass should be strictly zero. Searching for a rest mass of the photon
has therefore been one of the most important efforts on testing the validity of this assumption.

However, it is not possible to do any experiment that would firmly confirm the photon mass
is exactly zero. Considering an age of the Universe of about $10^{10}$ years, the ultimate
upper limit on the rest mass of the photon can be estimated to be $m_{\gamma}\approx\hbar/(\Delta t)c^2\approx10^{-66}~\rm{g}$
on the basis of the Heisenberg uncertainty principle, where $\hbar$ is the Plank constant \citep{2005RPPh...68...77T}.
Although such an infinitesimal mass would be hard to detect, there are some possible observable
effects associated with a nonzero photon mass for it. These effects include a frequency dependence
of the velocity of light in vacuum, deviations in the behaviour of static electromagnetic fields,
the existence of longitudinal electromagnetic radiation, the gravitational deflection of massive
photons, and so on. All these observable effects have been studied carefully and have been
applied to set upper limits on the photon mass, either through terrestrial/laboratory experiments
or astrophysical observations (see \citealt{1973PhRvD...8.2349L,2005RPPh...68...77T,2006AcPPB..37..565O,2010RvMP...82..939G,2011EPJD...61..531S}
for reviews).

The Particle Data Group \citep{2014ChPhC..38i0001O} suggests the currently accepted upper limit on
the photon mass is $m_{\gamma}\leq1.5\times10^{-51}~\rm{g}$, almost $10^{24}$ times
smaller than the electron mass, and this value was obtained by analyzing the magnetohydrodynamic
phenomena of the solar wind at Pluto's orbit \citep{2007PPCF...49..429R}. Since the photon mass
is quite impressively small, there is no essential effect on atomic and nuclear physics;
and it is extremely difficult to improve this limit by laboratory experiments.
However, even a very minor mass would has a significant effect on astrophysical phenomena
occurring at large distances over the photon Compton wavelength. This introduces the need
to develop methods as more as possible for constraining the photon mass with all kinds of
alternative astrophysical observations.

In astrophysical observations, the most direct and model-independent method for determining
the photon mass is to measure a frequency dependence of the velocity of light. Up to now,
limits on the photon mass through the dispersion of light have been made using the electromagnetic
emissions from flare stars \citep{1964Natur.202..377L}, Crab Nebula pulsar \citep{1969Natur.222..157W},
active galactic nuclei (AGN; \citealt{1999PhRvL..82.4964S}), high redshift Type Ia supernovae \citep{1999PhRvL..82.4964S},
gamma-ray bursts (GRBs; \citealt{1999PhRvL..82.4964S,2016JHEAp..11...20Z}), and fast radio bursts (FRBs;
\citealt{2016ApJ...822L..15W,2016PhLB..757..548B}). In particular, by analyzing the observed time delay between different
frequency photons from FRB 150418 at a cosmological distance $z=0.492$ \citep{2016Natur.530..453K},
\cite{2016ApJ...822L..15W} set the most strict limit to date on the frequency dependence of the speed of light,
implying a photon mass $m_{\gamma}\leq5.2\times10^{-47}~\rm{g}$. However, the redshift determination
of FRB 150418 has been questionable. \cite{2016ApJ...821L..22W} proposed that the so-called radio transient
of FRB 150418 may come from a common AGN variability and unrelated to FRB itself, and so
this redshift measurement may not be justified (but see \citealt{2016arXiv160304825L}). Although the cosmological origin of FRB 150418
is still under debate, \cite{2016ApJ...822L..15W} also showed that even if FRB 150418 originated in
our Local Group (1 Mpc), a stringent limit on the photon mass of $m_{\gamma} \leq 1.9\times10^{-45}~\rm{g}$
can be still obtained, which is already one order of magnitude better than the previous best result
from GRB 980703 ($m_{\gamma} \leq 4.2\times10^{-44}~\rm{g}$; \citealt{1999PhRvL..82.4964S}).

We note that although pulsar has been used to constrain the photon mass, this constraint relied on
the relative arrival time delay for pulses of the Crab Nebula pulsar over the optical wavelength
range of 0.35--0.55 $\rm \mu m$, yielding a limit of only $m_{\gamma}\leq5.2\times10^{-41}~\rm{g}$
\citep{1969Natur.222..157W}. Due to the fact that limits on the photon mass could be significantly improved
by several orders of magnitude with radio observations of pulsars at farther distances,
here we suggest that radio pulsars in the Large and Small Magellanic Clouds (LMC and SMC) are also
good candidates for constraining the photon mass. Compared to the prospects of the Crab Nebula pulsar
in constraining the photon mass, the radio pulsars in the LMC and SMC have two advantages. Firstly,
measurements of time structures at lower frequency are particularly powerful for constraining
the photon mass, and hence radio emissions of pulsars should be used for this purpose, rather than
optical emissions. Secondly, the distances of the LMC and SMC (49.7 and 59.7 kpc) are much more distant
than that of the Crab Nebula (2 kpc). The value of the distance also plays an important role in
constraining the photon mass, a larger distance leading to better constraints on the photon mass.
On the other hand, unlike FRBs would be subject to the distance uncertainty, radio pulsars
in the LMC and SMC have very certain distances, and the photon mass constraints from these pulsars
would be more reliable than those of FRBs.

In this work, we first try to constrain the photon rest mass using radio pulsars in the LMC and SMC.
The rest of the paper is organized as follows. In \S~\ref{sec:method}, we illustrate the velocity dispersion method
used for our analysis. New limits on the photon mass from radio pulsars are presented in \S~\ref{sec:result}.
Lastly, we summarize our conclusions in \S~\ref{sec:summary}.

\section{Description of the method}
\label{sec:method}
\subsection{Velocity dispersion from the nonzero photon mass}
According to Einstein's special relativity, the energy of the photon with a rest mass $m_{\gamma}$
can be expressed as
\begin{equation}\label{eq1}
E=h\nu=\sqrt{p^2c^2+m_\gamma^2c^4}\;.
\end{equation}
For the nonzero mass ($m_{\gamma}\neq0$) case, the speed of photon $\upsilon$ in vacuum is no longer a constant,
but depends on the frequency $\nu$. The dispersion relation is given by
\begin{equation}\label{eq2}
\upsilon=\frac{\partial{E}}{\partial{p}}=c\sqrt{1-\frac{m_\gamma^2c^4}{E^2}}=c\left(1-A\nu^{-2}\right)^{1/2}\approx c\left(1-\frac{1}{2}A\nu^{-2}\right)\;,
\end{equation}
where $A=m_\gamma^2c^4/h^2$.
One can see from Equation~(\ref{eq2}) that high frequency photons propagate in vacuum faster than
low frequency photons.

Consider two photons with different frequencies (denoted by $\nu_{h}$ and $\nu_{l}$,
where $\nu_{h}>\nu_{l}$) emitted simultaneously from a same source at distance $L$. Since the lower
frequency photon is slightly slower than the higher one, these two photons would be received at different times
by the observer, whose differences can be estimated as
\begin{equation}\label{eq3}
  \Delta{t_{m_{\gamma}\neq 0}}=\frac{L}{\upsilon_{l}}-\frac{L}{\upsilon_{h}}\approx\frac{LA}{2c}\left(\nu_l^{-2}-\nu_h^{-2}\right)\;.
\end{equation}
Thus, Equation~(\ref{eq3}) leads to constraints on the photon rest mass as
\begin{equation}\label{eq4}
 m_{\gamma}\approx hc^{-3/2}\left[\frac{2\Delta{t_{m_{\gamma}\neq 0}}}{\left(\nu_l^{-2}-\nu_h^{-2}\right)L}\right]^{1/2}\;.
\end{equation}

\subsection{Velocity dispersion from the plasma effect}
It is well known that the pulsar radiation would be affected by free electrons in the interstellar medium (ISM)
when they travel across the ISM, especially for the low energy radiation (e.g., the radio emissions; \citealt{2015aska.confE..41H,2015RAA....15.1629X}).
Due to the dispersive nature of plasma, or the ionized part of the atoms, the velocity dispersion of
photons would be related to
\begin{equation}\label{eq5}
\upsilon=c\left[1-\left(\frac{\nu_{p}}{\nu}\right)^{2}\right]^{1/2}\;,
\end{equation}
where $\nu_{p}=(ne^{2}/\pi m_{e})^{1/2}$ is the plasma frequency. Here $n$ is average electron density
along the line-of-sight, $m_{e}$ and $e$ is the mass and charge of the electron. Equation~(\ref{eq5})
also implies that higher frequency radio photons pass through the ISM faster than lower frequency ones
(see e.g., \citealt{2016arXiv160708820B}).

With Equation~(\ref{eq5}), we have the arrival time delay between two different
frequencies caused by the ISM plasma effect, i.e.,
\begin{equation}\label{eq6}
\begin{split}
  \Delta{t}=\frac{e^{2}}{2\pi m_{e}c}\left(\nu_l^{-2}-\nu_h^{-2}\right)\int n_{e}{\rm d}l\\
  =\frac{e^{2}}{2\pi m_{e}c}\left(\nu_l^{-2}-\nu_h^{-2}\right){\rm DM}~~~\;,
\end{split}
\end{equation}
where the dispersion measure (DM) represents the integrated electron density along the line-of-sight,
i.e., ${\rm DM}=\int n_{e}{\rm d}l$.

\subsection{Methodology}
Generally, the observed arrival time delay ($\Delta{t_{\rm obs}}$) for pulses of the pulsar between
different wavelengths is directly used to measure the DM. However, the observed time delay $\Delta{t_{\rm obs}}$
should be, in principle, mostly contributed by the nonzero photon mass ($m_{\gamma}\neq0$) effect
(if exists) and the plasma effect. In other words, both the massive photon and the line-of-sight
free electron content determine the same DM. Assuming that all of the DM measurement is dominated by
the $m_{\gamma}\neq0$ effect, a conservative upper limit on the photon mass can be estimated by
combining Equations~(\ref{eq3}) and (\ref{eq6}), i.e.,
\begin{equation}\label{eq:7}
m_{\gamma}\leq\frac{he}{c^{2}}\left(\frac{\rm DM}{\pi m_{e}L}\right)^{1/2}\;,
\end{equation}
which can be further reduced to
\begin{equation}\label{eq:8}
m_{\gamma}\leq\left(6.6\times10^{-45}\rm g\right)
\left(\frac{\rm DM}{\rm 100\; pc\; cm^{-3}}\right)^{1/2}\left(\frac{L}{\rm 10\; kpc}\right)^{-1/2}\;.
\end{equation}

\section{Photon mass limits from radio pulsars}
\label{sec:result}
The LMC and SMC are the closest galaxies to our Milky Way Galaxy and, so far, the only galaxies other than
our own that have detectable pulsars. Several surveys for radio pulsars in the Magellanic Clouds have been
carried out by \cite{1983Natur.303..307M}, \cite{1991MNRAS.249..654M}, \cite{2001ApJ...553..367C},
\cite{2006ApJ...649..235M}, and \cite{2013MNRAS.433..138R}, which discovered 21 radio pulsars in the LMC.
In addition, 5 radio pulsars were also discovered in the SMC in these surveys (see
\citealt{1991MNRAS.249..654M,2001ApJ...553..367C,2006ApJ...649..235M}), leading to a total of 26 radio pulsars
in the Magellanic Clouds. We now display how radio pulsars in the Magellanic Clouds provide an excellent way
of probing the photon mass by taking two typical extragalactic pulsars (PSRs J0451-67 and J0045-7042) as examples.
Here we choose these two pulsars with the smallest DM values in the LMC and SMC to get the best constraints on
the photon mass (see Equation~\ref{eq:8}).

\subsection{PSR J0451-67}
In a systematic survey of the LMC and SMC for radio pulsars, \cite{2006ApJ...649..235M} discovered 14 pulsars using the Parks
64-m radio telescope at 1400 MHz, 12 of which are believed to be in the Magellanic Clouds.
Of these 12 radio pulsars, nine in the LMC and three in the SMC. In addition, they appear to be usually located
in the central regions of each Cloud.

PSR J0451-67 is one of the LMC pulsars, with a mean flux density (averaged over the pulse period) $\sim0.05$ mJy
at 1400 MHz and the value of $\rm DM=45\;pc\; cm^{-3}$. Using the known distance ($L=49.7$ kpc) and DM
measurement of PSR J0451-67, a severe upper limit on the photon mass from Equation~(\ref{eq:8}) is
$m_{\gamma}\leq2.0\times10^{-45}\rm g$, which is 4 orders of magnitude tighter than that obtained by the Crab Nebula pulsar
\citep{1969Natur.222..157W}, and is as good as the result on FRB 150418 when the FRB originating within our Local Group
\citep{2016ApJ...822L..15W}.

\subsection{PSR J0045-7042}
In this extensive survey of \cite{2006ApJ...649..235M}, PSR J0045-7042 was found to lie within the SMC
($L=59.7$ kpc). It has a mean flux density $\sim0.11$ mJy at 1400 MHz, and its DM
measurement is $\rm 70\;pc\; cm^{-3}$. With these information of PSR J0045-7042, we can
restrict the photon mass from Equation~(\ref{eq:8}) to be $m_{\gamma}\leq2.3\times10^{-45}\rm g$,
which is also $10^{4}$ times better than the constraint from the Crab Nebula pulsar \citep{1969Natur.222..157W},
and is also as good as the result from FRB 150418 in our Local Group \citep{2016ApJ...822L..15W}.

\section{Conclusions}
\label{sec:summary}
The rest mass of the photon $m_{\gamma}$ can be effectively restricted by measuring the frequency dependence of
the speed of light. Using this dispersion method, we prove that radio pulsars in the Magellanic Clouds
can serve as a new excellent candidate for constraining the photon mass. Assuming that the whole DM
measurement of the pulsar between different radio bands is mainly due to the nonzero photon mass ($m_{\gamma}\neq0$) effect
and adopting the distance of the LMC or SMC, we place robust limits on the photon mass for two extragalactic pulsars:
$m_{\gamma}\leq2.0\times10^{-45}\rm g$ for the LMC pulsar PSR J0451-67 and $m_{\gamma}\leq2.3\times10^{-45}\rm g$
for the SMC pulsar PSR J0045-7042. Compared with the limit from the optical emissions of the Crab Nebula pulsar
\citep{1969Natur.222..157W}, our constraints on $m_{\gamma}$ with radio pulsars in the Magellanic Clouds
represent an improvement of at least 4 orders of magnitude.

Previously, the cosmological distance ($z=0.492$) of FRB 150418 provided the most stringent limit on
the photon mass through the dispersion method, showing an upper limit of $5.2\times10^{-47}~\rm{g}$
\citep{2016ApJ...822L..15W}. However, the redshift measurement of FRB 150418 remains controversial
(\citealt{2016ApJ...821L..22W}; but see \citealt{2016arXiv160304825L}), so the strict constraint on the photon mass from FRB 150418 may be not quite
reliable. But it is encouraging that even if FRB 150418 is not cosmological, the extragalactic
origin of FRB 150418 can still lead to a severe limit on the photon mass of $m_{\gamma} \leq 1.9\times10^{-45}~\rm{g}$
\citep{2016ApJ...822L..15W}. Although our limits on the photon mass ($\sim10^{-45}~\rm{g}$) are not as tight as the result of
the cosmological FRB, our limits can reach the high precision of the extragalactic FRB.

It should be underlined that our limits are based on a very conservative estimate of the DM measurement,
we suppose that all of the DM is mainly contributed by the $m_{\gamma}\neq0$ effect. In fact,
the DM measurement should be highly dominated by the plasma effect, with a very small contribution
possibly from the $m_{\gamma}\neq0$ effect. We find that if the $m_{\gamma}\neq0$ effect is responsible
for $10.0\%$ of DM of radio pulsars, much more severe limits could be obtained, implying
$m_{\gamma} \leq 6.3\times10^{-46}~\rm{g}$ for PSR J0451-67 and $m_{\gamma} \leq 7.3\times10^{-46}~\rm{g}$
for PSR J0045-7042.


\normalem
\begin{acknowledgements}
We are grateful to the anonymous referee and Martin Schlederer for helpful comments.
This work is partially supported by the National Basic Research Program (``973" Program)
of China (Grant Nos 2014CB845800 and 2013CB834900), the National Natural Science Foundation
of China (Grants Nos. 11322328, 11433009, 11673068, and 11603076), the Youth Innovation
Promotion Association (2011231), the Key Research Program of Frontier Sciences (QYZDB-SSW-SYS005),
the Strategic Priority Research Program ``The Emergence of Cosmological Structures"
(Grant No. XDB09000000) of the Chinese Academy of Sciences, the Natural Science Foundation
of Jiangsu Province (Grant No. BK20161096), and the Guangxi Key Laboratory for Relativistic Astrophysics.
\end{acknowledgements}


\end{document}